# Restricted thermodynamic fluctuations and the Ruppeiner geometry of black holes

Anurag Sahay[*]

*Department of Physics, National Institute of Technology Patna, Patna 800005, India*

Thermodynamic fluctuation metrics in Ruppeiner's formalism are worked out for Kerr-AdS black holes in the extended state space. The implications of constraints upon the state space geometry and their correspondence with thermodynamical ensembles are explicitly worked out in the most general setting. The state space scalar curvature for a given ensemble is found to be sensitive to the instabilities or phase transitions therein. In particular, it is found that the appropriate Ruppeiner scalar curvature does encode critical phenomena in the Kerr-AdS black holes. A detailed study is undertaken of the curvature contour of the state space of the 4D Kerr-AdS black hole, and suitable inferences are drawn. In particular, thermodynamic geometry suggests an instability in the Schwarzschild-AdS limit for all the ensembles except the pressure ensemble, which is equivalent to the unextended state space of the Kerr-AdS black holes. The extrinsic geometry of the ensemble hypersurfaces is introduced, and its relevance to constrained thermodynamic fluctuations is discussed. A new interpretation for the thermodynamic curvature of black hole systems is suggested.

## I. INTRODUCTION

It was first proposed by Bekenstein that black holes are genuine thermodynamic objects with an entropy which is a measure of all the information concerning the black hole interior that is lost to the outside world [1]. Drawing on the well-known analogy with the laws of black hole mechanics, he conjectured that the black hole entropy $S$ is a constant factor of order unity times the horizon area $A_H$ in Planck units. Subsequently, in a seminal work Hawking was able to show via a semiclassical calculation that black holes emit thermal radiation with a blackbody spectrum at a temperature proportional to their surface gravity, while the precise expression for the entropy worked out to $S = A_H/4$ in Planck units [2].

The large entropy of black holes implies a very large degeneracy $e^S$ of underlying microstates and has inspired a determined quest to seek its microscopic origin. Remarkable progress has been made in this endeavor by workers in quantum theories of gravity, especially for black hole solutions in string theory [3,4].[1] In particular, over the past decade the AdS/CFT correspondence, a remarkable realization of the gauge-gravity duality conjecture in string theory, has significantly enriched our perspective on black hole thermodynamics and phase transitions via the description of boundary field theory [6–8]. Thus, the Hawking-Page transition in the AdS black holes was shown to be the confinement-deconfinement phase transition in boundary gauge theory [9].

In a different context, pioneering efforts by Ruppeiner and others have shown that the equilibrium state space of a thermodynamic system can be uplifted to a Riemannian geometry via the Gaussian fluctuation moments which constitute the metric [10]. Thermodynamic geometry has been investigated for a variety of systems ranging from fluids and magnetic systems to various black hole systems, and significant information has been revealed through the invariants of the geometry, like the geodesics or the curvature scalar [10–15]. Thermodynamic geometry forms a remarkable connection from the thermodynamic description to the underlying statistical description. Thus, the magnitude of the scalar curvature invariant $R$ turns out to be proportional to the correlation volume of ordinary thermodynamic systems, while its sign is an indicator of the attractive or repulsive nature of the underlying statistical interactions. Further, it has been shown that apart from representing the critical point through its singularity, the scalar curvature also encodes first order phase transitions in simple fluids, uniquely determines the Widom line in different regimes, and identifies the experimentally determined solidlike patches in the liquid phases [16,17].

The program of black hole thermodynamic geometry aims to infer details of the underlying microscopic description of the black hole system starting from its thermodynamics. The geometrical view is important to pursue since, despite some remarkable achievements, the microscopic description of the thermodynamics and phase structure of black holes is a work in progress. Starting with the work of [18], in recent years there has been an upsurge of interest in the geometric approaches to black hole thermodynamics, most of them employing either the Ruppeiner (Weinhold) formalism, wherein the metric is the Hessian of the entropy (energy), or the more recent Quevedo "geometrothermodynamics" [19], which makes use of the contact structure of the phase space to obtain Legendre invariant thermodynamic metrics. Thermodynamic geometry has been investigated for various black holes in spacetimes with differing asymptotics and for different theories of gravity [20–51], where it has been shown that the state space scalar curvature $R$ is an effective probe for investigating instabilities and phase transitions. Of particular interest to a geometric analysis are the black holes in asymptotically anti–de Siter space (AdS black holes) with their dual description in terms of boundary gauge theory via the AdS/CFT duality. With the AdS space acting like a box, the canonical ensemble of AdS black holes is stable, as opposed to the asymptotically flat case, and their phase structure is analogous to ordinary systems like the van der Waals fluid [52,53]. In fact, the analogy with the phase structure of fluids becomes precise for the *extended* state space thermodynamics of AdS black holes, the subject of this work. In this scenario the cosmological constant $\Lambda$ is treated as a thermodynamic variable and acts like a pressure term, with its conjugate being the "thermodynamic volume" of the black hole [54–57]. The thermodynamics and phase behavior in the extended state space have been investigated recently [58–63], and interesting phenomena like reentrant phase transitions have been reported. The state space geometry of AdS black holes has been widely investigated, and recently the geometric analysis has been applied to dual gauge theory [64,65] and, in particular, to strongly coupled field theories like holographic superconductors [66]. The thermodynamic geometry of extended phase space has been investigated in [64,65,67–69].

---

[*]sahay.anurag12@gmail.com
[1]For an alternative viewpoint see [5].

The consensus seems to be that Ruppeiner's scalar curvature $R$ does not always encode all the phase transitions or instabilities in a given black hole system. Thus, for example, it is known that $R$ does not encode the van der Waals-like critical point in the Kerr-Newman-AdS black holes [45,46], and it was noticed in [67] that even the extended state space scalar curvature does not become singular at the critical points of the Kerr-Newman-AdS black holes. In [40–42] efforts made at generalizing Ruppeiner's metric by including the Hessian of thermodynamic potentials apart from entropy or energy resulted in scalar curvatures that captured different instabilities not encoded by $R$. While the alternative approaches are very promising, we nevertheless restrict ourselves to an analysis of the Ruppeiner geometry, which has a well-understood physical meaning from thermodynamic fluctuation theory. In general, therefore, it is not clear *a priori* whether or not the state space geometry is sensitive to the instabilities in a particular ensemble of the black hole system. We try to address this issue here. To this end we undertake a systematic study of the intrinsic and extrinsic geometry of hypersurfaces in the equilibrium state space and discuss their role in describing the thermodynamic processes and phase structures in a given ensemble. It turns out that phase transitions and instabilities in a given ensemble are captured by the geometry of the relevant hypersurface. We then explicitly work out Ruppeiner's geometry for different ensembles of the Kerr-AdS black hole in the extended state space. We report that, contrary to the prevailing viewpoint, the extended state space geometry corresponding to the fixed angular momentum case becomes singular at the van der Waals-like critical point of the 4D Kerr-AdS black hole. Therefore, Ruppeiner's geometry does encode second order phase transitions in the Kerr-AdS black hole. Furthermore, the geometry reveals an interesting "instability" in the extended state space in the Schwarzschild-AdS limit (or $J/M \to 0$) which is not captured by the usual response coefficients. Namely, for all the ensembles in which the fluctuations in the thermodynamic volume are treated as independent, the state space scalar curvature diverges to negative infinity in the zero angular momentum limit. Indeed, it is only for the unextended state space, namely, the constant pressure ensemble, that the geometry remains regular in the Schwarzschild AdS limit. This has to do with the fact that in the zero angular momentum limit, the thermodynamic volume is no longer an independent variable.

In this paper, we explicate the role of constrained fluctuations in determining codimension-one hypersurfaces in the (pseudo)Riemannian state space on which the black hole thermodynamic system "lives." Even though our approach is quite general in its applicability, we restrict our analysis to black hole systems and, in particular, to the extended thermodynamics of AdS black holes. Indeed, ensemble inequivalence of black hole systems helps us transparently illustrate the main features of our general approach. At the same time, the choice of the AdS black holes in the extended phase space reflects a growing interest in their thermodynamic properties and phase behavior as discussed above, and we hope that our analysis connects with efforts at placing them within the context of AdS/CFT correspondence [70].

Before we delve into the details of our proposal and its application to the Kerr-AdS black holes, we pause to sketch our perspective on the physical meaning of $R$ for black hole systems. Indeed, there is a general agreement that $R$ is a measure of interaction strength for the black hole thermodynamic system and that its divergence signals thermodynamic instability and phase transitions, just as for ordinary *extensive* systems. These general notions about the features of $R$ have been successfully tested for many black hole systems. However, it appears difficult to borrow the physical interpretation of $R$ from ordinary extensive thermodynamic systems for the case of black holes since for the latter no notion of a correlation length exists. This has to do with the nonextensivity of the black hole entropy (similar to other self-gravitating systems [71]), with the added complication that, in the absence of any information about its interior, the location of its microscopic degrees of freedom is not understood. Not surprisingly, most of the literature on the thermodynamic geometry of black holes has largely restricted its focus to the ability of $R$ to capture thermodynamic instabilities. An exception is the work by Ruppeiner [39,72–74], where he advances a quantum gravity inspired proposition that $R$ is a measure of the number of correlated Planck-sized pixels at the horizon of the black hole.

In an upcoming work we discuss an alternative interpretation of $R$ which appears to be especially suited to the black hole context and probably to other nonextensive statistical systems as well [75]. We sketch it very briefly here. Thus, for extensive systems, where the entropy is an order one homogeneous function of extensive variables, it is well known that $R_{\text{extn}}$ has units of volume, and that $R_{\text{extn}} \sim \xi^d$, the correlation volume. Here, the subscript under $R$ is for extensive. Moreover, it relates to the singular part of the free entropy density, namely, $R_{\text{extn}} \sim \psi_s^{-1}$. Now, especially near the critical point, the singular part of the free entropy *counts* the number of organized fluctuations in the system, $\Psi_s = L_0^d \psi_s \sim L_0^d / \xi^d$, and goes to zero at the critical point. Therefore, a nondimensionless form of thermodynamic curvature suggests itself, namely, $R' = R_{\text{extn}}/L_0^d \sim \Psi_s^{-1}$, where $d$ is the spatial dimension of the system. Therefore, this has the straightforward interpretation that $1/R'$ counts the number of correlation volumes, or the number of statistically independent *domains* in the system. Now, as can be checked, near the critical point the nondimensionless scalar curvature $R$ of the black hole scales as the inverse of the singular part of the free entropy [46]. Thus, based on the above we could think of the curvature length scale $l_R^2 = 1/|R|$ of the black hole system as a count of the number of statistically independent domains. Of course, in the black hole context, by domain we mean a collection of correlated degrees of freedom for which we do not know the distribution or location in space. Evidently, this approach is similar in spirit to, but slightly different in emphasis from, the aforementioned Ruppeiner's approach. We defer the details of our proposal to our forthcoming work [75], where we also discuss thermodynamic curvature in an AdS/CFT framework.

This paper is organized in two sections, as follows. In Sec. II and its subsections we establish the connection between thermodynamical ensembles, which we carefully define, and the corresponding hypersurfaces living in the ambient state space. In Sec. II A we present a picture of the Ruppeiner's metric as a Mahalanobis norm for state space fluctuations about equilibrium. In Sec. II B we work out the thermodynamic metrics corresponding to restricted state space fluctuations in a given thermodynamic ensemble. Finally, in Sec. II C we discuss the extrinsic and intrinsic geometry of the ensemble hypersurfaces. In Sec. III and its subsections, we extensively apply the ideas developed in the first part to explore the geometry of the extended state space of 4D Kerr-AdS black holes. In Sec. IV we conclude our discussion with brief comments and some key points.

For clarity and completeness, we sketch the development of our ideas starting from the basic ingredients of thermodynamic fluctuation geometry and differential geometry, which to many might appear elementary in places. We take that risk nevertheless since we feel that an understanding of the program of geometrization of (restricted) fluctuations requires us to engage with a few nuances at the heart of the issue.

## II. GEOMETRY OF CONSTRAINED FLUCTUATIONS

In this section we work out our method for obtaining the thermodynamic geometry corresponding to restricted fluctuations in the state space.

### A. Thermodynamic fluctuation metrics: A view from scatter plot statistics

In the following we briefly review the setup of Ruppeiner's thermodynamic geometry before presenting an active view of the state space Riemannian metric as a measure of the spontaneous motion of the system in the state space manifold around its equilibrium point. This alternative viewpoint complements the static view of the metric as a distance measure between nearby probability distributions.

The thermodynamic fluctuation theory for black holes envisages the black hole system and the reservoir in a mutual equilibrium, which together form a microcanonical system with a conserved and additive set, $X_0 = X + X_r$, of charges with the exception of the entropy, $S_{\text{tot}} = S + S_r$, which is additive but not conserved. Unlike the black hole, the reservoir is an extensive system and is weakly coupled to the black hole so that their respective charges remain additive [27,28]. The thermodynamic state of the reservoir, which is fixed by definition and is labeled by its (entropic) intensive variables or "potentials" $\theta_r$, fixes the equilibrium point $\bar{X}$ of the black hole in its equilibrium state space.[2] At any other point $X$ in its state space, while the black hole remains in an internal equilibrium; it is no longer in mutual equilibrium with the reservoir $\theta_r$. The probability distribution of a spontaneous fluctuation of the black hole state from the above mentioned equilibrium point $\bar{X}$ to a point $X$ in its state space is determined by the entropy of the overall microcanonical system,

$$p(X;\theta)d^n X = C \exp[S_{\text{tot}}(X;\theta)]d^n X \qquad (1)$$

---

[2]Tuning the reservoir potentials is formally equivalent to equilibrating the black hole with a succession of reservoirs whose states $\theta_r$ differ incrementally from each other [76].

where

$$S_{\text{tot}}(X;\theta) = S(X) + S_{r,\theta}(X_0 - X) \leq S_{\text{tot}}(\bar{X};\theta). \quad (2)$$

On Taylor expanding the total entropy around its maximum value at equilibrium, we get

$$S_{\text{tot}}(X;\theta) = S_{\text{tot}}(\bar{X};\theta) + \frac{1}{2}d^2 S + \cdots. \quad (3)$$

The second term in the expansion above is the second order change in the black hole entropy, and it must be negative for a stable equilibrium between the black hole and the reservoir,

$$d^2 S = \frac{\partial^2 S}{\partial X^\mu \partial X^\nu}(\bar{X}^\mu - X^\mu)(\bar{X}^\nu - X^\nu) = -g_{\mu\nu}\Delta X^\mu \Delta X^\nu \quad (4)$$

where the indices $\mu$, $\nu$ run across all the extensive charges.

In the small fluctuation approximation only the quadratic term is retained, and the normalized probability distribution becomes

$$P(\bar{X} + \Delta X)d^n X = \frac{1}{(2\pi)^{n/2}}\exp\left[-\frac{1}{2}g_{\mu\nu}\Delta X^\mu \Delta X^\nu\right] \\ \times \sqrt{g(\bar{X})}d^n X, \quad (5)$$

where $g(\bar{X})$ is the determinant of the "metric" $g_{\mu\nu}$ at $\bar{X}$.

For a stable equilibrium $g_{\mu\nu}$ is positive definite and transforms as a covariant tensor of second rank under a general coordinate transformation $X^\mu \to X'^\mu$, while $\Delta X^\mu$'s transform contravariantly, as can be easily checked [10]. This renders the expression inside the exponential invariant under coordinate changes and supplies a natural definition of an invariant distance between equilibrium points. Unlike the full probability distribution in Eq. (1), the Gaussian approximation above becomes covariant since, besides the line element, the expression outside the exponential is invariant, too.

The Massieu function $\Psi$, or the free entropy, is obtained from the entropy via a Legendre transformation as[3]

$$\Psi = S - \theta_\mu X^\mu \quad (6)$$

where the entropic intensive variables are $\theta_\mu = \partial S/\partial X^\mu$ and their infinitesimal changes $d\theta_\mu$ transform covariantly,

$$-d\theta_\mu = -\frac{\partial \theta_\mu}{\partial X^\nu}dX^\nu = g_{\mu\nu}dX^\nu. \quad (7)$$

The extensive variables are similarly obtained from the Massieu function as $X^\mu = -\partial \Psi/\partial \theta_\mu$. The first law of thermodynamics can be equivalently expressed by infinitesimal changes in $S$ or $Ψ$,

$$dS = \theta_\mu dX^\mu \quad \text{and} \quad d\Psi = -X^\mu d\theta_\mu. \quad (8)$$

The second order change in $\Psi$ and that in $S$ are of opposite sign and give the thermodynamic line element,

$$d^2\Psi = -d^2 S = -d\theta_\mu dX^\mu. \quad (9)$$

Equation (9) above is a useful starting point for expressing the line element in terms of the available black hole parameters. It is straightforward to check that the thermodynamic metric

$$g_{\mu\nu} = -\frac{\partial^2 S}{\partial X^\mu \partial X^\nu} \quad (10)$$

has its inverse in the Hessian of $\Psi$,

$$g^{\mu\nu} = \frac{\partial^2 \Psi}{\partial \theta_\mu \partial \theta_\nu}. \quad (11)$$

Thus,

$$g^{\alpha\nu}g_{\nu\beta} = \frac{\partial X^\nu}{\partial \theta_\alpha}\frac{\partial \theta_\beta}{\partial X^\nu} \\ = \delta^\alpha{}_\beta. \quad (12)$$

The metric and its inverse have a clear meaning in terms of the second moment of thermodynamic fluctuations. The inverse metric equals

$$g^{\mu\nu} = \langle \Delta X^\mu \Delta X^\nu \rangle, \quad (13)$$

while the Hessian metric equals

$$g_{\mu\nu} = \langle \Delta \theta_\mu \Delta \theta_\nu \rangle. \quad (14)$$

In this paper we do not privilege the extensive variables over the intensive ones as regards fluctuations, in as much as both are thermodynamic properties *of* the black hole. In some treatments an asymmetry is introduced between the extensive and intensive variables so that the latter represent the potentials of the reservoir, which has a fixed thermodynamic state. In the present work, however, following Einstein's fluctuation theory the microcanonical intensive variables $\theta = \partial S/\partial X$ fluctuate about their mean value $\theta_r$ fixed by the reservoir [10]. Simply put, in a given ensemble, any of the parameters of the black hole—like the horizon radius $r_h$, the angular momentum parameter $a$, the charge parameter $q$, or even the AdS length scale $l$ in the extended state space scenario—might fluctuate about their given values. Therefore, all the quantities that are dependent on them, whether the mass or the temperature, would undergo ensemble-dependent spontaneous fluctuations. From Eq. (14) above we can calculate the second fluctuation moments of intensive variables like $T$, $\Phi$ etc.

In the following, while we discuss the entropic metric in (10) as the "Hessian metric," we sometimes refer to its inverse in Eq. (11) as the "covariance metric" $\boldsymbol{\Sigma}$ for obvious reasons. The covariance metric relates directly to the information geometry of equilibrium thermodynamics and is the Fisher-Rao metric on the statistical manifold parametrized by the variables $\theta$. Thus, the line element between two nearby equilibrium points in the state space is equivalently interpreted as the distance measure between two nearby probability distributions in the parameter manifold. In Fig. 1 above we indicate the equilibrium state space of the black hole as $\mathbf{E}$. We also depict the parameter space $\mathbf{M}$ in the figure wherein each point $\theta$ labels the state of the reservoir and hence fixes a point $\bar{X}$ in $\mathbf{E}$ where the black hole equilibrates with the reservoir at $\theta$.[4] More precisely, a point $\theta_1$ in the parameter space induces a unique probability distribution $p(X; \theta_1)$ on the state space $\mathbf{E}$ given by Eq. (1) which could informally be viewed as a "scatter plot" with its "center" at $\bar{X}_1$. Similarly, $\theta_2$ induces a scatter plot on $\mathbf{E}$ determined by $p(X; \theta_2)$ and centered around $\bar{X}_2$, etc. Therefore, a state space point $X_1$ could either be the center $\bar{X}_1$ of a fixed distribution $p(X; \theta_1)$ or serve as a random sample vector for any other distribution, say, $p(X; \theta_2)$. This reflects in the two related interpretations of the Hessian matrix, Eq. (10), at a point on the state space. As discussed above, on the one hand, it provides a distance measure from the center $\overline{X_1}$ to a nearby center $\overline{X_2}$, or, equivalently, between two probability distributions $\theta_1$ and $\theta_2$, thus rendering the state space into a Riemannian manifold, $\{\mathbf{E}, g_{\mu\nu}\} \to \mathcal{E}$, or, equivalently, the parameter space into a parameter manifold, $\{\mathbf{M}, g^{\mu\nu}\} \to \mathcal{M}$. At the same time, given a probability $p(X; \theta_1)$ on $\mathbf{E}$ centered around the mean $\bar{X}_1$, the matrix, Eq. (10), is the Mahalanobis norm [77] for any other state space "sample vector" $\mathbf{X} - \bar{\mathbf{X}}$, assuming that the scatter is Gaussian. A sample-statistic definition of the Mahalanobis distance in the context of state space is given as

$$d_M^2(\mathbf{X}, \bar{\mathbf{X}}) = (\mathbf{X} - \bar{\mathbf{X}})^T \mathbf{\Sigma}^{-1} (\mathbf{X} - \bar{\mathbf{X}}) \qquad (15)$$

where $\mathbf{\Sigma}^{-1}$ is simply the Hessian metric, Eq. (10). It is a scale-invariant distance measure and counts the number of (multivariate) standard deviations a sample point happens to be from the mean. For a state space vector close to the mean in Eq. ([77]), the Mahalanobis distance is identical to the line element in Eq. (4), where $\Delta X^\mu$ is a tangent space vector.

---

[3] Note hthat the extensive and intensive variables, $X^\mu$ and $\theta_\nu$, respectively, are not themselves vector quantities. The suggestive placement of indices against these could be confusing but should be kept in mind [10].

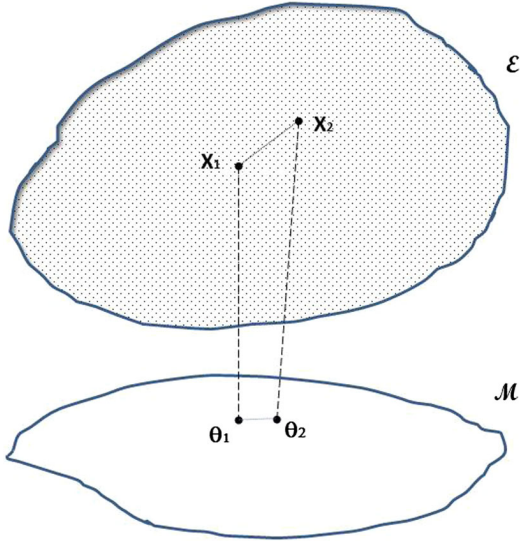

FIG. 1. The state space manifold along with the parameter manifold. Each point on the parameter manifold $\mathcal{M}$ fixes an equilibrium point on $\mathcal{E}$ about which there is a distribution of state space fluctuations.

---

[4] A one-to-one mapping between the set $X$ and its conjugate set $\theta$ exists as long as the Jacobian matrix, Eq. (7), between the two is invertible, implying that the system is locally stable. At the same time, two locally stable equilibrium branches could coexist for a range of values of $\theta$, in which case Fig. 1 refers to a specific branch.

Therefore, informally, one could picture the tangent space $T_X$ associated with a point $X$ in the state space manifold $\mathcal{E}$ as being spanned by the spontaneous fluctuation vectors $\Delta X^\mu$ around the equilibrium state at $X$. This could be thought of as a kinematic view of the state space wherein the thermodynamic system continuously scans the tangent space around its equilibrium point $\bar{X}_1$ with the probability distribution function $P(\bar{X}_1 + X)$, Eq. (5). In this respect, the foregoing discussion simply (semi)formalizes our intuition for the state space metric. The eigenvectors associated with the covariance metric specify the directions of statistically independent equilibrium fluctuations around a state space point, and the square root of its eigenvalues is the variance along the respective independent directions. As a result, the square root of its determinant is proportional to the "covariance ellipsoid" around a point and could be thought of as a measure of the statistical distinguishability of two nearby distributions.

### B. Fluctuations, constraints, and the ensemble hypersurface

In the following we set up the connection between ensembles and hypersurfaces. In order to fix some of the notions to be developed in this section, we take as our reference black hole system the asymptotically AdS 4D Kerr-Newman (KN-AdS) black hole. This is the most general black hole solution of Einstein-Maxwell's equation in four dimensions and is characterized by its ADM mass and angular momentum, $M$ and $J$, respectively, and its electric charge $Q$. In the following general discussion, we do not require the exact Smarr formula for our black hole, and it suffices that there exists a fundamental relation for the entropy in terms of extensive variables, $S \equiv S(M, Q, J)$. The three-dimensional (unextended) state space of the KN-AdS black hole may be described by any three independent coordinates as long as the Jacobian of transformation is nonzero, which will be the case in a locally stable region.

Thus, for example, the set $\Omega, T, \Phi$ could equivalently serve as the state space coordinates.

Any given path in the state space manifold, considered as a set of equilibrium points, has fluctuations around all its points in the sense discussed in the last subsection. Now, along a given path, if some variables, like $Q$ or $\Omega$, are held to constant values, such curves will lie on hypersurfaces of constant $Q$, $\Omega$, etc. in the state space manifold. However, the constancy of $Q$ or $\Omega$ does not preclude spontaneous equilibrium fluctuations in their values. At a point on the hypersurface, whether the equilibrium fluctuations are restricted to the tangent space of the hypersurface or they extend to the full tangent space of the embedding manifold is to be determined by the nature of the constraint imposed.[5] Physically, it could be that fluctuations in some modes of the black hole system are very slow compared to others, in which case one could entirely neglect the slower modes as a first approximation.

A general case is one where the path lies entirely on a hypersurface given by some function,

$$f(E, Q, J) = f_0. \quad (16)$$

If that function is, say, the electric potential $\Phi$, then the path corresponds to an isopotential process. Once again, the nature of the constraint determines whether the potential $\Phi$ is strictly constant or constant in the mean along the isopotential path. We define an $f$ constraint to be a strict restriction to an $f$ hypersurface of the overall state space motion of the black hole system which includes its equilibrium fluctuations.

The set of allowed fluctuations therefore determines the nature of the ambient state space in which the path exists, or, in other words, it determines the ensemble in which the thermodynamic process takes place. In this respect, thermodynamical ensembles are shorthand for the quantities we treat as thermodynamic variables as opposed to the ones we fix as parameters [78]. In this paper's terminology we reserve the term grand canonical for the ensemble wherein there is no restriction on the fluctuations in all the state space variables and the black hole is in full thermodynamic contact with the reservoir with all its charges freely participating in the exchange. On the other hand, when there is an $f$ constraint present, as defined above, then we refer to the ensemble as $f$ canonical. Thus, if the angular momentum $J$ of the KN-AdS black hole is held strictly constant, then it is referred to as being in the $J$-canonical ensemble, or, for brevity, simply the $J$-ensemble. Similarly, it could be in an $\Omega$-ensemble, and so forth. The most restrictive case in which the only mode of energy exchange is in the form of heat will be referred to simply as the canonical ensemble.

We now discuss the relation between thermodynamic fluctuation metrics and ensembles. In Eq. (3) the entropy of a nearby fluctuation was obtained via a Taylor expansion around the equilibrium state without any constraint imposed on the charges. Such unconstrained fluctuations in all the charges of a system are a defining feature of the grand canonical ensemble, as discussed above. Thus, for $N$ extensive charges the grand canonical metric, or the "grand" metric, would be $N$ dimensional. In the presence of a constraint the thermodynamic metric is no longer $N$ dimensional. As discussed earlier, an $f$ constraint restricts the thermodynamic motion of a black hole system, including its spontaneous fluctuations, to a codimension-one hypersurface in its state space. The metric corresponding to the restricted fluctuations is naturally the one induced on the $f$ hypersurface from the grand metric. We term such a metric the $f$ canonical metric, or simply the $f$-metric.

Before we discuss some geometrical aspects of ensemble hypersurfaces in the next subsection, we briefly implement the discussion above in the context of the KN-AdS black hole. The infinitesimal change in the KN-AdS black hole entropy is given by the first law of thermodynamics,

$$dS = \frac{1}{T}dE - \frac{\Phi}{T}dQ - \frac{\Omega}{T}dJ, \quad (17)$$

and the line element between nearby equilibrium states is obtained as

$$-dl^2 = d\theta_\mu dX^\mu = d\left(\frac{1}{T}\right)dE + d\left(-\frac{\Phi}{T}\right)dQ + d\left(-\frac{\Omega}{T}\right)dJ. \quad (18)$$

The inverse grand metric is obtained from the following Massieu function,

$$\Psi = S - \frac{E}{T} + \frac{\Phi}{T}Q + \frac{\Omega}{T}J. \quad (19)$$

In the presence of constraints the full three-dimensional grand metric will induce a metric on the relevant hypersurface. For most of the physically meaningful constraints, the induced metric can be obtained from the grand metric by a direct observation. Thus, for the ensemble in which one or more of the charges (or extensive variables) are constrained, the induced metric is simply obtained by dropping off the terms containing the variation of those charges in Eq. (18). For a fixed angular momentum hypersurface, or the $J$-ensemble, for example, the induced metric is obtained by a simple truncation of the line element of Eq. (18),

$$-dl_J^2 = d\left(\frac{1}{T}\right)dE + d\left(-\frac{\Phi}{T}\right)dQ. \quad (20)$$

---

[5]Note that compared to the intensive variables, it might appear more natural to constrain the extensive variables, or the charges, by means of "walls." However, we shall not concern ourselves with such questions.

The relevant Massieu function is obtained by the partial Legendre transform of the entropy,

$$\Psi_J\left(\frac{1}{T}, -\frac{\Phi}{T}\right) = S - \frac{1}{T}E + \frac{\Phi}{T}Q. \quad (21)$$

We could also freeze the fluctuations in an intensive variable like the angular velocity so that on the $\Omega$ hypersurface the variations in $J, Q, E$ are related by the condition $\delta\Omega(J, Q, E) = 0$, thus reducing the number of unconstrained statistical variations to two. The metric for this $\Omega$ ensemble is similarly written down by an observation from Eq. (18),

$$-dl_\Omega^2 = d\left(\frac{1}{T}\right)d(E - \Omega J) + d\left(-\frac{\Phi}{T}\right)dQ. \quad (22)$$

A suitable choice of extensive variables is suggested from the form of the metric above, namely, $W = E - \Omega J$ and $Q$, so that the first law is rewritten as

$$dS = \frac{1}{T}dW + \frac{\Phi}{T}dQ. \quad (23)$$

Thus, the black hole could be thought of as exchanging the independent extensive variables $W$ and $Q$ with the reservoir. The fluctuations in the $\Omega$ ensemble give rise to the covariance metric which is obtained from the following Massieu function,

$$\Psi_\Omega\left(\frac{1}{T}, \frac{-\Phi}{T}\right) = S - \frac{1}{T}W + \frac{\Phi}{T}Q. \quad (24)$$

The metric induced on the $\Omega$ surface now records fluctuations in $W$ and $Q$,

$$h_\Omega^{ij} = \langle \Delta Y^i \Delta Y^j \rangle, \quad (25)$$

where $\{Y^1, Y^2\} \equiv \{W, Q\}$.

We note that our prescription for obtaining the geometries associated with restricted fluctuations could be viewed as a generalization of Ruppeiner's approach in [27].

### C. Extrinsic and intrinsic geometry of the ensemble hypersurface

To fix our discussion we could think of the KN-AdS black hole once again. Thus, in the three-dimensional state space of the KN-AdS black hole, let there be a hypersurface given by Eq. (16). The unit normal vector to the $f$ hypersurface is proportional to its gradient and has the standard expression

$$n_\mu = \epsilon \frac{\partial_\mu f}{\sqrt{|g^{\mu\nu}\partial_\mu f \partial_\nu f|}} \quad (26)$$

where $g^{\mu\nu}$ is the grand metric of the state space while $\epsilon$ is the sign of the norm of $n_\mu$ and is negative when it is "timelike." If the normal vector $n^\mu$ is timelike at a point, then that point is unstable to fluctuations along $n^\mu$ since such a variation results in an increase in the entropy. Therefore, the suppression of fluctuations along $n^\mu$ carves out a slice of stability in a state space region which is otherwise unstable to unrestricted fluctuations.

We now briefly discuss a geometrical method of obtaining constrained fluctuation moments in a given ensemble alluded to earlier. The metric on the hypersurface is induced from the ambient metric by projecting out the directions associated with the normal vector,

$$g_{(f)}^{\mu\nu} = g^{\mu\nu} - \epsilon n^\mu n^\nu. \quad (27)$$

The "projection metric" introduced above is the same as the induced metric discussed previously. In the current representation the indices run over the full set of state space coordinates as opposed to the hypersurface specific representation wherein the constraints have been explicitly taken into account. The two are of course related as follows,

$$h_{(f)ij} = g_{(f)\mu\nu}\frac{\partial X^\mu}{\partial Y^i}\frac{\partial X^\nu}{\partial Y^j} \quad (28)$$

where the $Y^i$'s are coordinates specific to the hypersurface.

In a straightforward manner, the projection metric represents the second order moments of the constrained fluctuations in the entropic intensive variables and the charges,

$$g_{(f)\mu\nu} = \langle \Delta\theta_\mu \theta_\nu \rangle_f \quad (29)$$

and

$$g_{(f)}^{\mu\nu} = \langle \Delta X^\mu \Delta X^\nu \rangle_f. \quad (30)$$

Keeping in mind the $\Omega$ ensemble of the KNAdS black hole in order to fix the ideas, we find that the projection metric in Eq. (27) and the induced metric in Eq. (25) play complementary roles in describing fluctuations. Thus, the $Q$ fluctuations in the $\Omega$ ensemble are equally well described by the induced metric and the projection metric through their respective components:

$$\langle (\Delta Q)^2 \rangle_\Omega = h_{(\Omega)}^{QQ} \equiv g_{(\Omega)}^{QQ} = g^{QQ} - \epsilon n^Q n^Q \quad (31)$$

where the normal vector is orthogonal to the $\Omega$ hypersurface. On the other hand, the constrained fluctuations in $M$ and $J$ are naturally described via the projection metric,

$$\langle (\Delta M)^2 \rangle_\Omega = g_{(\Omega)}^{MM}, \quad (32)$$

$$\langle (\Delta J)^2 \rangle_\Omega = g_{(\Omega)}^{JJ}, \quad (33)$$

and

$$\langle \Delta M \Delta J \rangle_\Omega = g^{MJ}_{(\Omega)}. \tag{34}$$

The induced metric Eq. (25) gives the constant $\Omega$ fluctuations in the quantity $W = M - \Omega J$ mentioned earlier,

$$\langle (\Delta W)^2 \rangle_\Omega = h^{WW}_{(\Omega)} \tag{35}$$

which can be related to the constrained fluctuations in $M, J$ via the projection metric,

$$\langle (\Delta W)^2 \rangle_\Omega = \langle (\Delta M)^2 \rangle_\Omega + \Omega^2 \langle (\Delta J)^2 \rangle_\Omega - 2\Omega \langle \Delta M \Delta J \rangle_\Omega. \tag{36}$$

Thus we see that the geometrical prescription offers a direct means to calculate fluctuations in the presence of very general constraints. Finally, we extend the aforementioned method to include fluctuations in the presence of multiple constraints. Thus, let $f_1(X^\mu) = C_1$ and $f_2(X^\mu) = C_2$ be two constraints present in $s$ state space of dimension greater than 2. Then the constrained fluctuations in the charges $X^\mu$ can be represented as

$$\langle \Delta X^\mu \Delta X^\nu \rangle_{f_1, f_2} = g^{\mu\nu} - \epsilon_1 n^\mu_{(1)} n^\nu_{(1)} - \epsilon_2 n^\mu_{(12)} n^\nu_{(12)}$$
$$= g^{\mu\nu}_{(1)} - \epsilon_2 n^\mu_{(12)} n^\nu_{(12)} \tag{37}$$

where

$$n^\mu_{(1)} = \epsilon_1 \frac{g^{\mu\nu} \partial_\nu f_1}{\sqrt{|g^{\alpha\beta} \partial_\alpha f_1 \partial_\beta f_1|}}, \tag{38}$$

while

$$n^\mu_{(12)} = \epsilon_2 \frac{g^{\mu\nu}_{(1)} \partial_\nu f_2}{\sqrt{|g^{\alpha\beta}_{(1)} \partial_\alpha f_2 \partial_\beta f_2|}}, \tag{39}$$

with $\epsilon_1$ and $\epsilon_2$ being the signatures of the norms of $n_{(1)}$ and $n_{(12)}$ in the metrics $g_{\mu\nu}$ and $g_{(1)\mu\nu}$, respectively.

When the total thermodynamic motion of the black hole system, including its spontaneous fluctuations, is restricted to an $f$ hypersurface, then, expectedly, the projection upon the hypersurface of the full scalar curvature $R$ gives its intrinsic scalar curvature $R_f$. Now the scalar curvature $R_f$ obtained from the induced metric $h_{(f)ij}$ is related to the grand curvature $R$ via the Gauss-Codazzi relation

$$R^{(3)} = R^{(2)}_f + \epsilon(n^\mu_{;\mu} n^\nu_{;\nu} - n^\mu_{;\nu} n^\nu_{;\mu}) + 2\epsilon(n^\mu_{;\nu} n^\nu - n^\mu n^\nu_{;\nu})_{;\mu}$$
$$= R^{(2)}_f + \epsilon(K^2 - K_{\mu\nu} K^{\mu\nu}) + 2\epsilon(n^\mu_{;\nu} n^\nu - n^\mu n^\nu_{;\nu})_{;\mu}.$$

Note that, just for this equation, we use superscripts on $R$ and $R_f$ to explicitly indicate their dimensions. Here, the covariant derivative is with respect to the grand metric $g_{\mu\nu}$, and the quantity $K = n^\mu_{;\mu}$ is the trace of the extrinsic curvature $K_{\mu\nu}$ which is the projection upon the hypersurface of the derivative $n_{\mu;\nu}$. The extrinsic curvature, or the second fundamental form, gives information about the embedding of the hypersurface in the ambient state space and is written as [79]

$$K^{\mu\nu} = g^{\mu\alpha}_{(f)} g^{\nu\beta}_{(f)} n_{\alpha;\beta}$$
$$= \frac{1}{2} g^{\mu\alpha}_{(f)} g^{\nu\beta}_{(f)} \mathcal{L}_n g_{\alpha\beta}$$
$$= \frac{1}{2} \mathcal{L}_n g^{\mu\nu}_{(f)}, \tag{40}$$

where the last equality can be interpreted, using the identity in Eq. (30), as the Lie derivative of the constrained fluctuations. It is generally the case that the intrinsic curvature $R_f$ of the $f$ hypersurface might become singular at a point but not the *grand* scalar curvature $R$ of the ambient space at that point and vice versa. Thus, for example, at the critical point of the $Q$ ensemble in the KN-AdS black hole, while the curvature $R_Q$ diverges the full curvature $R$ remains regular [45,46]. The latter would therefore remain insensitive to the $Q$-ensemble instabilities. We stress this point here since it has often led to confusion in the literature.

Recently, a work appeared that is related to extrinsic curvatures in state space [80].

## III. ADS BLACK HOLES AND EXTENDED STATE SPACE THERMODYNAMICS

Recently, there has been active interest in exploring the outcome of treating the cosmological constant of black hole spacetimes as a thermodynamic variable [58–64,67–69]. It was demonstrated in [54] by utilizing Komar integrals and Killing potentials in static AdS spacetimes, that $\Theta$, the conjugate to $\Lambda$ in the first law of thermodynamics, has the dimensions of volume. In fact, the quantity $\Theta$ was demonstrated to be the negative of the volume excluded by the black hole in spacetime. This fit well with the natural interpretation of $\Lambda$ as a pressure term. As a result, the ADM mass of $M$ of the black hole was reinterpreted as the enthalpy of the black hole, while its internal energy $E$ was obtainable as a Legendre transform of the enthalpy. Thus, for the four-dimensional spacetime

$$E(S, V, Q, J) = M(S, P, Q, J) - PV, \tag{41}$$

where the pressure is related to $\Lambda$ as $P = -\Lambda/8\pi$ and the thermodynamic volume is given as $V = -8\pi\Theta$, where Newton's constant $G_4$ has been set to unity. The geometrical derivation was further completed and elaborated upon for more general charged and rotating black holes in [55] for a negative cosmological constant. While for static solutions the thermodynamic volume matched the naive interpretation of the volume as an integral from the singularity to the horizon, for the rotating case there was an additional term proportional to the angular momenta.

In this work we deal with extended state space for black holes in the anti–de Sitter spacetime. The first law of thermodynamics for the KN-AdS black hole is now given as

$$dS = \frac{1}{T}dE + \frac{P}{T}dV - \frac{\Omega}{T}dJ - \frac{\Phi}{T}dQ, \quad (42)$$

while the line element in the extended state space is

$$-dl^2 = d\left(\frac{1}{T}\right)dE + d\left(\frac{P}{T}\right)dV$$
$$+ d\left(-\frac{\Phi}{T}\right)dQ + d\left(-\frac{\Omega}{T}\right)dJ \quad (43)$$

where $E$ refers to the internal energy of the black hole from Eq. (41). The thermodynamic volume $V$ is now an extensive variable which can be exchanged with the environment. Thus, $V$, as well as $P$, is now subject to thermodynamic fluctuations around their mean equilibrium values. To recall our discussion relating to the fluctuations, the pressure $P$ could be held constant in two ways, either strictly or in the mean. The former case gives the nonextended thermodynamics wherein we have frozen out the $P$ fluctuations. More explicitly, we have

$$P(E, V, Q, J) = P_0 \quad (44)$$

so we are strictly restricted to the $P$ surface in the four-dimensional extended state space. In this $P$ ensemble the number of independent variables is reduced to three because of the above constraint. The induced line element can be obtained from the grand line element, Eq. (43), by a simple rearrangement,

$$-dl_P^2 = d\left(\frac{1}{T}\right)d(E + PV)$$
$$+ d\left(-\frac{\Phi}{T}\right)dQ + d\left(-\frac{\Omega}{T}\right)dJ \quad (45)$$

where now the three independent extensive variables that the black hole exchanges with the environment are conveniently given by $Q$, $J$ and the enthalpy $M = E + PV$. The first law, Eq. (42), is similarly modified.

It turns out that the extended state space of the RN-AdS black hole is not that useful from the point of view of thermodynamic geometry. This is because the thermodynamic volume of the RN-AdS black hole, as for any static AdS spacetime, is a monotonic function of the entropy and hence does not undergo an independent fluctuation.[6] While we could address the more general case of charged rotating AdS black holes in arbitrary dimensions, for the sake of simplicity, we restrict ourselves to the extended state space geometries belonging to the 4D Kerr-AdS black hole since it already exhibits interesting critical phenomena.

---

[6]At the same time this property of $V$ for static AdS spacetimes turns out to be very useful in the realization of holographic heat engines [62].

### A. Extended fluctuation geometries for the 4D-Kerr AdS black holes: The grand ensemble

The 4D Kerr-AdS black hole is described by three independent parameters in the Boyer-Lindquist coordinate frame. These are the horizon radius $r$, the angular momentum parameter $a$, and the AdS length scale $l$ which is related to the pressure via the relation $|\Lambda| = 3/l^2$. Note that in this article we denote the horizon radius by $r$ instead of the standard notation $r_h$. In terms of these parameters the extensive variables in the extended state space are given as

$$M = \frac{J}{a} = \frac{l^2(a^2 + r^2)(l^2 + r^2)}{2(a-l)^2(a+l)^2 r}, \quad (46)$$

$$S = \frac{\pi(a^2 + r^2)}{1 - \frac{a^2}{l^2}}, \quad (47)$$

$$V = \left(\frac{\partial M}{\partial P}\right)_{S,J} = \frac{2l^2\pi(a^2 + r^2)(a^2 l^2 - a^2 r^2 + 2l^2 r^2)}{3(a-l)^2(a+l)^2 r},$$
$$\quad (48)$$

and

$$E = M - PV = \frac{(a^2 + r^2)(2l^4 a +^2 r^2 - a^2 l^2)}{4(a-l)^2(a+l)^2 r}, \quad (49)$$

while the intensive variables are

$$T = \frac{a^2(-l^2 + r^2) + r^2(l^2 + 3r^2)}{4l^2 \pi r(a^2 + r^2)}, \quad (50)$$

$$P = \frac{3}{8\pi l^2}, \quad (51)$$

and

$$\Omega = \frac{a(l^2 + r^2)}{l^2(a^2 + r^2)}. \quad (52)$$

Note that the angular velocity $\Omega$ above is the difference between the horizon angular velocity $\Omega_H$ and the velocity of the Boyer-Lindquist frame at infinity, which is $\Omega_\infty = -a/l^2$. It is the same as the velocity of the rotating Einstein universe at the boundary of AdS space on which the dual gauge theory lives as per the AdS/CFT correspondence [53]. For angular velocity $\Omega < 1/l$ the black hole can be in equilibrium with a corotating heat bath filling up the AdS space up to infinity, while for $\Omega > 1/l$ the black hole develops superradiant instabilities and the Einstein universe spins faster than light.

The temperature goes to zero at the zero of the extremal polynomial,

$$\mathcal{N}_{ex} = a^2(-l^2 + r^2) + r^2(l^2 + 3r^2). \quad (53)$$

From the expression for entropy, it is clear that $a \leq l$.

The extended first law of thermodynamics in this ensemble includes the variations in all the extensive variables,

$$dS = \frac{1}{T}dE + \frac{P}{T}dV - \frac{\Omega}{T}dJ. \quad (54)$$

The grand metric $g_{\mu\nu}$ is given by the line element

$$-dl^2 = d\left(\frac{1}{T}\right)dE + d\left(\frac{P}{T}\right)dV + d\left(-\frac{\Omega}{T}\right)dJ. \quad (55)$$

In this paper we order the state space coordinates as $\{X^1, X^2, X^3\} \equiv \{E, V, J\}$ and $\{\theta_1, \theta_2, \theta_3\} \equiv \{1/T, P/T, -\Omega/T\}$.

The covariance metric can be obtained via the following grand Massieu potential: $\Psi(1/T, P/T, -\Omega/T) =$

$$-\frac{l^2\pi(l^2 - r^2)(a^2 + r^2)^2}{(a^2 - l^2)(a^2(l^2 - r^2) - r^2(l^2 + 3r^2))} \quad (56)$$

with its determinant $\mathcal{D}$ given as

$$\mathcal{D} = \frac{a^4 l^2 (l^2 + r^2)(a^2(l^2 - r^2) - r^2(l^2 + 3r^2))^5}{288(a^2 - l^2)^6 \pi r^6 (r^2(l^2 - 3r^2) + a^2(l^2 + r^2))}. \quad (57)$$

Note that the determinant $\mathcal{D}$ vanishes in the limit where $a$ goes to zero. The reason the covariance metric becomes degenerate in this limit is because, for $a = 0$, the thermodynamic volume $V$ becomes a function of $E$ and hence its fluctuations are no longer independent of $E$.

The maximum number of independent thermodynamic response functions relevant to the ensemble can be neatly obtained from the six second partial derivatives of the Gibbs free energy with respect to its natural parameters,

$$G(T, P, \Omega) = -T\Psi = E - TS + PV - \Omega J. \quad (58)$$

They are the heat capacity $C_{P\Omega} = T(\frac{\partial S}{\partial T})_{P\Omega}$,

$$C_{P\Omega} = \frac{2l^2\pi r^2(a^2(-l^2 + r^2) + r^2(l^2 + 3r^2))}{(a^2 - l^2)(r^2(l^2 - 3r^2) + a^2(l^2 + r^2))}, \quad (59)$$

the expansivity $\alpha_{P\Omega} = \frac{1}{V}(\frac{\partial V}{\partial T})_{P\Omega} =$

$$-\frac{4l^2\pi r(6l^2 r^4 + 3a^2 r^2(l^2 - r^2) + a^4(l^2 + r^2))}{(2l^2 r^2 + a^2(l^2 - r^2))(r^2(l^2 - 3r^2) + a^2(l^2 + r^2))}, \quad (60)$$

$$\left(\frac{\partial J}{\partial T}\right)_{P\Omega} = -\frac{2al^4\pi(l^2 + r^2)(a^4 + 4a^2 r^2 + 3r^4)}{(a^2 - l^2)^2(r^2(l^2 - 3r^2) + a^2(l^2 + r^2))}, \quad (61)$$

the compressibility $\kappa_{T\Omega} = -\frac{1}{V}(\frac{\partial V}{\partial P})_{T,\Omega} =$

$$\frac{16l^2\pi(9l^4 r^6 + 9a^2 l^2 r^4(l^2 - r^2) + a^6(l^4 - r^4) + a^4(5l^4 r^2 - l^2 r^4 + 3r^6))}{3(a^2 - l^2)(2l^2 r^2 + a^2(l^2 - r^2))(r^2(l^2 - 3r^2) + a^2(l^2 + r^2))}, \quad (62)$$

the moment of inertia $\mathcal{I}_{PT} = (\frac{\partial J}{\partial \Omega})_{P,T} =$

$$\frac{l^4(a^2 + r^2)(-l^4 r^4 + 3l^2 r^6 + a^6(l^2 + r^2) + a^4(3l^4 + 13l^2 r^2 + 6r^4) + a^2(6l^4 r^2 + 23l^2 r^4 + 9r^6))}{2(a^2 - l^2)^3 r(r^2(l^2 - 3r^2) + a^2(l^2 + r^2))}, \quad (63)$$

and

$$\left(\frac{\partial J}{\partial P}\right)_{T\Omega} = -\frac{4al^4\pi(a^2 + r^2)(l^2 + r^2)(9l^2 r^4 + a^4(2l^2 + r^2) + a^2(7l^2 r^2 - 3r^4))}{3(a^2 - l^2)^3 r(r^2(l^2 - 3r^2) + a^2(l^2 + r^2))}. \quad (64)$$

These functions carry the same information as the grand metric, and any other response function can be obtained from these six using the Maxwell relations [81]. Expectedly, the divergence of these response functions follows that of the covariance determinant in Eq. (57) above. In the state space parametrized by the variables $r$, $l$, and $a$, the divergence occurs on the hypersurface, which is the zero of the following polynomial:

$$\mathcal{N} = r^2(l^2 - 3r^2) + a^2(l^2 + r^2) = 0. \quad (65)$$

We name this hypersurface the Davies surface. Following the previous discussion, in this ensemble all the extensive variables $E$, $V$, and $J$ of the black hole fluctuate around their mean values fixed by the environment. The second moments of their fluctuations are given by the corresponding components of $g^{\mu\nu}$ as discussed previously. Similarly, the second moments of the standard intensive quantities, namely, $T$, $P$, and $\Omega$, can be obtained from the components of $g_{\mu\nu}$ which produce the moments of the $\theta_i$'s. We list the simple relations between variations of the entropic intensive variables and those of the standard intensive ones,

$$\Delta T = -\frac{1}{\theta_1^2}\Delta\theta_1, \tag{66}$$

$$\Delta P = \frac{1}{\theta_1}\Delta\theta_2 - \frac{\theta_2}{\theta_1^2}\Delta\theta_1, \tag{67}$$

$$\Delta\Omega = -\frac{1}{\theta_1}\Delta\theta_3 + \frac{\theta_3}{\theta_1^2}\Delta\theta_1. \tag{68}$$

These equations, in conjunction with Eq. (14), can be used to obtain the second moments of fluctuations in $P$, $T$, etc. Thus, for example, the second moment of $\Omega$ fluctuations may be obtained from above as

$$\langle(\Delta\Omega)^2\rangle = \theta_1^{-2}g_{33} + \theta_3^2\theta_1^{-4}g_{11} - 2\theta_3\theta_1^{-3}g_{13}$$
$$= \frac{(a^2 - l^2)^2 \mathcal{N}_{ex}}{2l^6\pi(a^2 + r^2)^3}. \tag{69}$$

Similarly, the fluctuations in $T$ and $P$ can be obtained as

$$\langle(\Delta T)^2\rangle = \frac{(l^2 - a^2)(18a^2l^4r^4 + 9l^4r^6 + a^6(l^4 - r^4) + a^4(5l^4r^2 - 10l^2r^4 - 6r^6))\mathcal{N}_{ex}}{8a^4l^6\pi^3r^2(a^2 + r^2)^3(l^2 + r^2)} \tag{70}$$

and

$$\langle(\Delta P)^2\rangle = \frac{9(l^2 - a^2)^2\mathcal{N}_{ex}}{32a^4l^4\pi^3(a^2 + r^2)^2(l^2 + r^2)}. \tag{71}$$

Interestingly, while the $\Omega$ fluctuations in the grand ensemble remain finite in the Schwarzschild-AdS limit, $a = J/M = 0$, the $T$ and $P$ fluctuations diverge in the same limit. The above equations indicate that there is a price for maintaining the volume fluctuation as an independent mode in the limit $J/M \to 0$. The price is an increase in the spontaneous fluctuations of the surface gravity on the black hole horizon. Note that the aforementioned six response coefficients which represent standard deviations of extensive variables are all finite in this limit. However, presumably, higher moments would be needed to describe the distribution of fluctuations where its temperature and pressure are strongly fluctuating. We expect the thermodynamic curvature $R$ to be able to describe the situation since it is an invariant measure of the validity of the Gaussian approximation. In fact, as we will see in the following, it clearly does so.

The state space scalar curvature in the grand ensemble was calculated in [67], and we express it in our coordinates below:

$$R = -\frac{\mathcal{P}_1}{a^4l^2\pi(a^2 + r^2)^2(l^2 + r^2)^2}\frac{1}{\mathcal{N}^2\mathcal{N}_{ex}} \tag{72}$$

where the polynomial $\mathcal{P}_1$ is given in the Appendix [Eq. (A1)].

Clearly, the grand scalar curvature diverges to negative infinity along the Davies surface, $\mathcal{N}$, and hence encodes the Davies phase transition in the grand ensemble. It also diverges to positive infinity at extremality, as usual for all black hole systems. Furthermore, $R$ becomes singular in the limit $a$ goes to zero. This is expected from the behavior of the determinant $\mathcal{D}$ in Eq. (57) since the state space geometry degenerates to a two-dimensional one at $a = 0$. This also connects with the diverging fluctuations in $T$ and $P$ in the $a = 0$ limit as pointed out in Eqs. (70) and (71) and the subsequent discussion. Therefore, it appears that thermodynamic geometry confirms the fact that when fluctuations in $V$ are independent of $E$ and $J$, as in the grand ensemble of the extended state space here, the $a = 0$ limit is untenable, and moreover, it becomes more and more difficult to sustain the grand ensemble for small $a$. We discuss this more towards the end of this section.

In Fig. 2 we plot, in the three-dimensional state space, the blue colored Davies surface, with $R$ diverging to negative infinity on it. The red colored surface is the extremal surface at which $T = 0$ and $R$ diverges to positive infinity on it. As far as the AdS black holes are concerned, in this work we focus mostly on the locally stable regions of the state space. Indeed, this is not to say that the geometry of the locally unstable regions is ill defined. For instance, the asymptotically flat black holes have negative heat capacity and a well-defined thermodynamic curvature [12]. In the case of black hole systems, a local instability, as reflected in the wrong sign of some response functions, might not be as catastrophic as for an ordinary extensive system. In the latter case a negative specific heat, for example, would mean that its parts become unstable to exchange of heat *amongst* themselves causing a thermal runaway within the system. For the black hole system a negative heat capacity does not cause an internal disruption other than rendering it unstable in the (grand) canonical ensembles, which implicitly assume the system to be in contact with an infinite bath. Furthermore, the statement that the wrong sign of the response functions implies a negative mean squared fluctuation of extensive charges and hence is a no-go for even the existence of, say, negative specific heat states is true only for extensive systems, which the black holes are not [71]. In any case, the geometry of such regions helps one identify trends of instability and slices of stability in them, as we will see in the following.

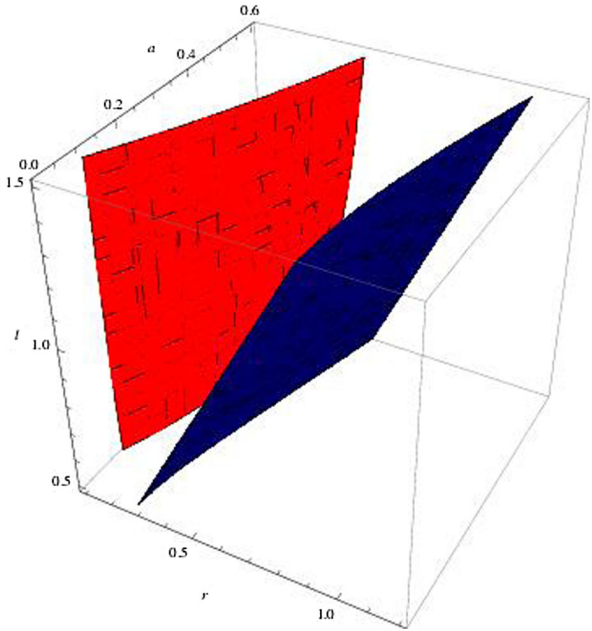

FIG. 2. Plot showing the hypersurface of the *R* singularity in the state space parametrized by *r*, *a*, and *l*. The blue colored singular surface, on which *R* has a negative divergence, partitions the state space in the grand canonical ensemble into a locally thermodynamically stable region below it and a locally unstable region behind it. The red colored singular surface is the extremal surface at which $T = 0$ and *R* has a positive divergence.

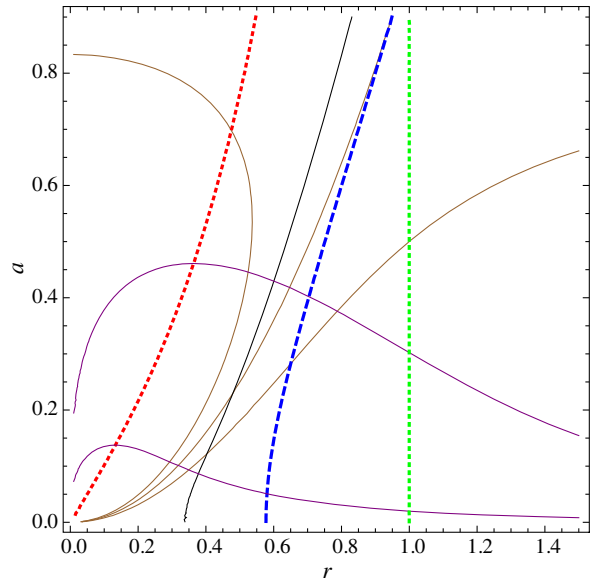

FIG. 3. A projection of Fig. 2 on a constant *P* surface at $l = 1$, with the dotted red and dotted blue curves being, respectively, the intersection of the extremal surface and the surface of *R*'s negative divergence. *R* is zero on the black curve and negative to its right. Gibbs free energy is zero on the green dotted curve and negative to the right. Some thermodynamic processes that are shown are purple colored constant *J* curves ($J = 0.02, 0.4$ from below to above) and brown colored constant $\Omega$ curves ($\Omega = 1.2, 1, 0.8$ from left to right).

Viewed geometrically, the state space is equipped with a thermal metric of varying signature. The scalar curvature in a specified ensemble diverges whenever one of the eigenvalues of the ensemble metric becomes zero. The locus of all degenerate points in the state space is thus a set of surfaces separating regions with a metric of different signature. In the full state space geometry corresponding to the grand canonical ensemble, the 3D metric has a signature (3,0) in the stable region, while in the unstable region it has a signature of (2,1) throughout, up to zero temperature, and hence is timelike there. Across the Davies surface one of the eigenvalues of the metric changes sign by passing through zero, while the other two remain unaffected. Across the extremal surface and into the forbidden negative temperature region, the metric has signature (1,2). In Fig. 3 we show some standard thermodynamic curves on a constant *P* slice of the state space. Note that the curves $\Omega \geq 1$ always lie in the thermodynamically unstable region of the grand ensemble. Also note the change in sign of *R* from negative to positive at low temperatures. We remind ourselves that in the grand ensemble, at all points on each of the curves, there exist fluctuations in every possible direction, including the ones away from the *P* surface. It is only to the right of the dotted blue surface that the equilibrium states become fully locally stable under any fluctuation. Finally, global stability *vis-à-vis* AdS space is achieved only to the right of the dotted green curve, which marks the zero of the Gibbs free energy.

We now make the observation that when moving away from the singular Davies surface into the locally stable part of state space, the grand curvature *R* does not always diminish. In Fig. 4 we plot *R* against the horizon radius *r* for a fixed *J* and *P*. Note that *J* and *P* are fixed only in the mean since in the grand ensemble all fluctuations are unconstrained. Clearly, as opposed to general intuition, *R* does not decay to zero everywhere in the locally stable region. While *R* decays to zero in the northeast direction, in Fig. 3 it grows to large negative values in the southeast direction where the geometry becomes singular at the $a = 0$ edge of this region. A more detailed picture of the landscape is offered by Fig. 5, which plots contours of *R* in the *r*-*a* plane with $l = 1$. From the figure we infer that while *R* decays along curves of constant *T* and $\Omega$, it increases in magnitude, after it briefly falls away from the Davies line, along curves of constant *E*, *J*, and *V*.

As mentioned earlier, *R* is not the only measure of curvature in the 3D state space manifold of the Kerr-AdS discussed in previously. We defer a detailed discussion of the implications of the full curvature tensor to a future work.

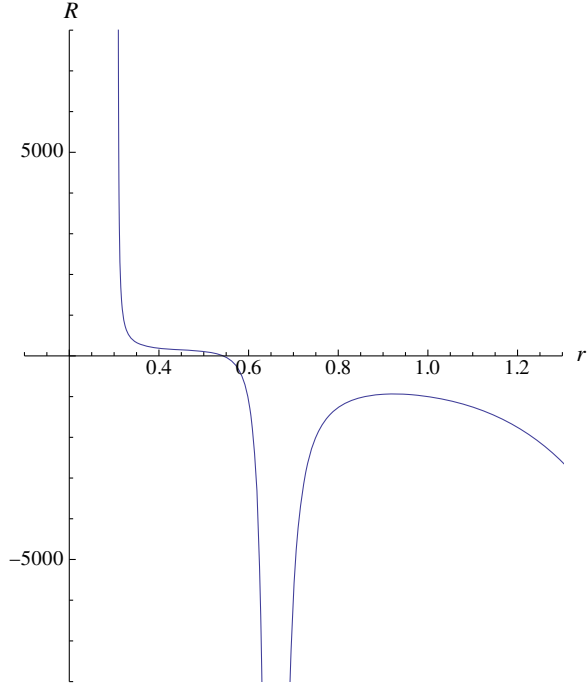

FIG. 4. A plot of the grand scalar curvature $R$ vs the horizon radius $r$ in the grand ensemble along a path with $J = 0.2$ and $l = 1$. The left arm corresponds to the locally unstable phase. Along the locally stable right arm, $R$ rises again after climbing down from a divergence at the Davies point.

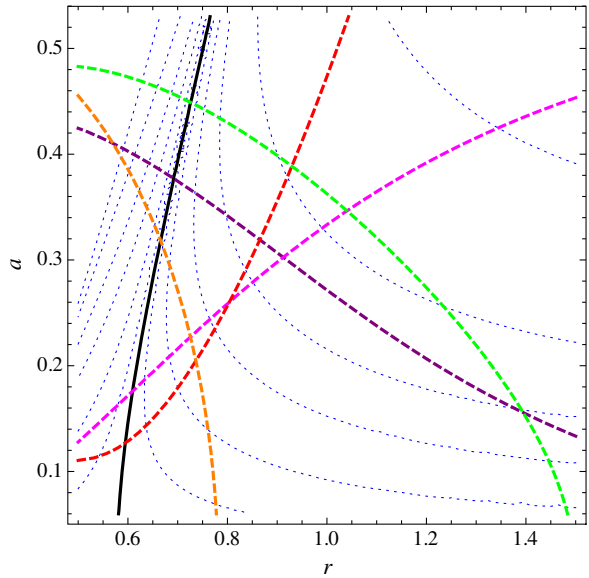

FIG. 5. A plot of contours of the grand scalar curvature $R$, shown as the dotted blue lines, in the constant $P$ plane. The thick black curve is the Davies curve where $R$ is singular. The dashed red and magenta curves have, respectively, $T$ and $\Omega$ constant, while the dashed green, purple, and orange curves have, respectively, $E$, $J$, and $V$ constant.

### B. The $J$ canonical ensemble: Critical phenomena

In this ensemble the angular momentum is fixed and its fluctuations are frozen, so the black hole's thermodynamic motion is strictly restricted to a $J$ hypersurface in its three-dimensional state space. The Kerr-AdS black hole exhibits a rich phase structure in the $J$ ensemble, with phase transitions of both first order and second order occurring between the small and large black holes, with critical exponents known to have mean field values [45,53,67]. The phase structure is further enriched in the higher-dimensional Kerr-AdS black holes, which display reentrant phase transitions, among other interesting phenomena [58,60]. As explained in [78] the region of stability increases with constraints, so the locally stable region of the grand canonical ensemble is always a subset of that for any canonical ensemble. Following the discussion in the previous section, there is a simple geometrical way to understand this. Thus, we stated earlier that for the grand ensemble, the region of instability (behind the blue colored Davies surface in Fig. 2) is characterized by a negative sign of any one of the eigenvalues of the grand metric in Eq. (55), which has a signature of (2,1) there. Indeed, while any hypersurface inherits all the locally stable regions of the ambient state space, it could continue to remain stable even in the locally unstable region of the ambient space if the unstable direction of the grand metric there remains parallel to the hypersurface normal. Thus, the ensemble hypersurfaces can carve out slices of stability in regions which are otherwise unstable under the full set of fluctuations. At the same time, the unstable eigendirection of the grand metric need not align with the hypersurface normal everywhere, so the ensemble hypersurface will become locally unstable too wherever the two directions are not parallel.

For the $J$ ensemble the full line element, Eq. (55), reduces to

$$-dl_J^2 = d\left(\frac{1}{T}\right)dE + d\left(\frac{P}{T}\right)dV, \quad (73)$$

while the Massieu function becomes

$$\Psi_J(1/T, P/T) = S - \frac{1}{T}E - \frac{P}{T}V. \quad (74)$$

This gives rise to a two-dimensional covariance metric whose determinant is

$$\mathcal{D}_J = \frac{a^4(l^2 + r^2)(\mathcal{N}_{ex})^4}{36(l^2 - a^2)^3 r^4 \mathcal{N}_J}, \quad (75)$$

where the zeros of the polynomial $\mathcal{N}_J$ determine the hypersurface on which the determinant $\mathcal{D}_J$ diverges,

$$\mathcal{N}_J = (-l^4 r^4 + 3l^2 r^6 + a^6(l^2 + r^2)$$
$$+ a^4(3l^4 + 13l^2 r^2 + 6r^4)$$
$$+ a^2(6l^4 r^2 + 23l^2 r^4 + 9r^6)). \quad (76)$$

Note that, similar to the grand determinant $\mathcal{D}$ in Eq. (57), the determinant $\mathcal{D}_J$ goes to zero in the Schwarzschild-AdS limit $a = 0$. The independent response functions in this ensemble, namely, $C_{JP}$, $\kappa_{JT}$, and $\alpha_{JP}$, are obtained from the three second partial derivatives of the Gibbs free energy $G_J(T, P) = -T\Psi_J$. They are as follows:

$$C_{JP} = \frac{2l^4 \pi (a^2 + r^2)^2 \mathcal{N}_{ex}}{(l^2 - a^2)\mathcal{N}_J}, \quad (77)$$

$$\kappa_{TJ} = \frac{16l^2 \pi (a^2 + r^2)(18a^2 l^4 r^4 + 9l^4 r^6 + a^6(l^4 - r^4) + a^4(5l^4 r^2 - 10l^2 r^4 - 6r^6))}{3(2l^2 r^2 + a^2(l^2 - r^2))\mathcal{N}_J}, \quad (78)$$

and

$$\alpha_{JP} = \frac{(4l^2(a^2 - l^2)\pi r(a^2 + r^2)(6l^2 r^4 + a^4(l^2 + r^2) + 3a^2(3l^2 r^2 + r^4)))}{(2l^2 r^2 + a^2(l^2 - r^2))\mathcal{N}_J}. \quad (79)$$

The normal vector to the $J$ surface points along the direction of increasing angular momentum and is hence easily expressed in the standard state space coordinates, which represent the extensive variables. Thus,

$$n_{(J)\mu} = \text{Sign}(g^{33})\frac{\delta_{3\mu}}{\sqrt{|g^{33}|}},$$
$$n^{\mu}_{(J)} = \text{Sign}(g^{33})\frac{g^{3\mu}}{\sqrt{|g^{33}|}}. \quad (80)$$

The normal vector $n_{(J)}$ can be used to construct the projection metric as in Eq. (27). Given that $J$ is a coordinate function on the state space, the contravariant form of the projection metric $g^{\mu\nu}_{(J)}$ does not supply any additional information compared to the explicitly two-dimensional induced metric from Eq. (73). However, in its covariant form it does provide additional information about constrained fluctuations in the intensive variables. For example, the microcanonical fluctuations in $\Omega$ on a $J$ surface can be evaluated via a modification of Eq. (69),

$$\langle (\Delta\Omega)^2 \rangle_J = \theta_1^{-2} g_{(J)33} + \theta_3^2 \theta_1^{-4} g_{(J)11} - 2\theta_3 \theta_1^{-3} g_{(J)13}$$
$$= \frac{2a^2(a^2 - l^2)^2(l^2 + r^2)(2l^2 r^2 + 3r^4 + a^2(l^2 + 2r^2))\mathcal{N}_{ex}}{l^6 \pi (a^2 + r^2)^3 \mathcal{N}_J} \quad (81)$$

where

$$g_{(J)\mu\nu} = g_{\mu\nu} - \epsilon n_{(J)\mu} n_{(J)\nu}. \quad (82)$$

Note that unlike the case of the grand ensemble in Eq. (69), the $\Omega$ fluctuations for the $J$ ensemble approach zero in the AdS-Schwarzschild limit $a = 0$. Moreover, since the $\Omega$ fluctuations diverge along the spinodal line, it becomes increasingly meaningless to talk about angular velocity in the critical region. At the same time, this does not reflect any difficulty in thermodynamic progress as alluded to earlier since, in the $J$ ensemble, there is no fixed $\Omega_r$ of the reservoir as a reference. The fluctuations in $T$ and $P$, given by Eqs. (70) and (71), respectively, are finite in the critical region but diverge in the limit $a = 0$. Therefore, the $J$ ensemble becomes increasingly difficult to sustain in the Schwarzschild-AdS limit, too. We expect that the interesting thermodynamics in the $J$ ensemble would be reflected by the scalar curvature, which we now discuss.

The scalar curvature in the $J$ ensemble is found to be

$$R_J = \frac{\mathcal{P}_2}{a^4 \pi (l^2 + r^2)^2 (\mathcal{N}_{ex})(\mathcal{N}_J)^2} \quad (83)$$

where the polynomial $\mathcal{P}_2$ is given in the Appendix [Eq. (A2)]. Evidently, $R_J$ encodes the thermodynamic instabilities of the $J$ ensemble, including its critical point, as can be seen from the presence of the polynomial $\mathcal{N}_J$ in its denominator. This is significant since, contrary to the current opinion, we see that Ruppeiner's geometry does encode critical phenomena in Kerr-AdS black holes.

We reemphasize that even though it is defined everywhere in the state space, $R_J$ is not a scalar curvature of the three-dimensional ambient space but an intrinsic curvature of the $J$-hypersurface family. Therefore, for a given thermodynamic process represented by a curve in the $J$ surface, $R_J$ will not necessarily be the correct invariant. It is only when the $J$ fluctuations are completely suppressed in the thermodynamic process that its instabilities are represented by $R_J$. Similar to the grand curvature $R$ in Eq. (72), the $J$ ensemble curvature $R_J$ also diverges in the $a = 0$ limit. Therefore, once again, geometry informs us that it is increasingly difficult to evolve a thermodynamic process, say, along an isotherm in the $J$-canonical ensemble of the extended state space once the ratio $J/M$ becomes small.

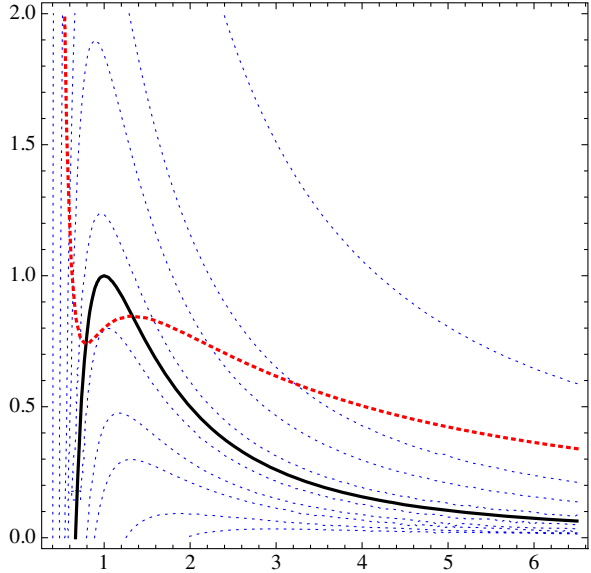

FIG. 6. A plot of contours of $R$ for the van der Waals gas in the $P$-$v$ plane. We use expressions from [16] for the universal vdW equation. The red dotted line is an isotherm.

In Figs. 6 and 7, respectively, we plot the $P - v$ and $P - r$ curves for the van der Waals (vdW) gas and the $J$

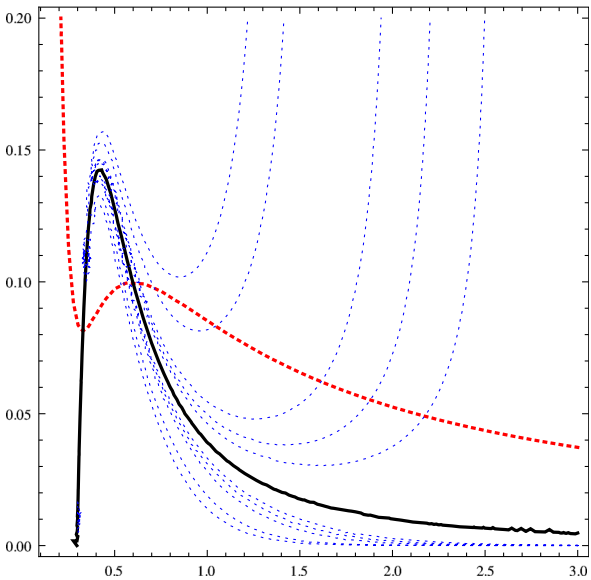

FIG. 7. A plot of contours of $R_J$ in the $P$-$r$ plane on the $j = 0.02$ hypersurface. Note the difference in contours of scalar curvature.

ensemble, respectively. The similarities of the extended phase space structure of the Kerr-AdS black hole with the vdW gas are well known, as can be observed from the two adjacent figures. Thus, similar to the vdW case the black hole is characterized by the coexistence of the small and large black hole phases which culminates in a critical point that has mean field exponents for both. However, there is a noticeable difference between the two state space geometries in the way their respective scalar curvatures fall off away from the spinodal line. For the vdW gas the fall off is standard, implying that the fluctuations become more and more Gaussian away from the spinodal line. On the other hand, in accordance with the discussion in the preceding paragraph, $R_J$ tends to grow, after a brief fall off, away from the spinodal line along standard curves, like the isotherm represented in Fig. 7.

The scalar curvature $R_J$ may also be obtained from the grand curvature $R$ via the Gauss-Codazzi relations, Eq. (40), which highlights aspects of embedding of the $J$ surface in the 3D state space. The trace $K_J$ of the extrinsic curvature of the $J$ surface is found to be

$$K_J = \text{Sign}(g^{33}) \frac{\mathcal{P}_3(l^2 - a^2)^{\frac{1}{2}}}{al(a^2 + r^2)(l^2 + r^2)} \frac{N_{ex}}{NN_J} \left| \frac{2N}{\pi N_{ex} N_J} \right|^{\frac{1}{2}} \quad (84)$$

where the polynomial $\mathcal{P}_3$ is expressed in the Appendix [Eq. (A3)].

From the expression above we see that $K_J$ encodes the instabilities intrinsic to the ensemble hypersurface as well as those of the ambient space. Furthermore, it vanishes at extremality. We defer a detailed study of the extrinsic curvature to a future investigation.

### C. The $P$ canonical ensemble

In this ensemble the fluctuations in the pressure $P$ are frozen out, so it corresponds to a strict restriction of the black hole thermodynamic motion to a constant $P$ hypersurface in its state space,

$$P(E, V, J) = P_0, \quad (85)$$

as we discussed previously in Eq. (45). The $P$ ensemble gives the unextended phase space thermodynamics of the Kerr-AdS black hole, with its natural choice of extensive variables being $\{Y^1, Y^2\} = \{M, J\}$. Indeed, the $P$ ensemble corresponds to the grand canonical ensemble of unextended state space of the Kerr-AdS black holes, and its thermodynamics, phase structure, and thermodynamic geometry have been thoroughly discussed in many works [45,46,53]. In this section we approach the state space geometry using the new insights developed in the present work. The metric induced on the ensemble gives the line element

$$-dl_P^2 = -h_{(P)ij}dY^i dY^j = d\left(\frac{1}{T}\right)dM - d\left(\frac{\Omega}{T}\right)dJ. \quad (86)$$

The inverse metric $h_{(P)}^{\mu\nu}$, i.e., the covariance metric, can be obtained from the second partial derivatives of the Massieu function,

$$\Psi_P(1/T, -\Omega/T) = S - \frac{M}{T} + \frac{\Omega J}{T} \quad (87)$$

with its determinant given as

$$\mathcal{D}_P = \frac{l^2 \mathcal{N}_{ex}^4}{64(a^2 - l^2)^4 \pi^2 r^4 \mathcal{N}}. \quad (88)$$

Unlike the previous two cases the determinant remains finite in the limit of $a$ going to zero. The three independent response functions obtained as second partial derivatives of the free energy $G_P(T, \Omega) = -T\Psi_P$ are $C_{P\Omega}$, $\mathcal{I}_{PT}$, and $(\partial J/\partial T)_{P\Omega}$ obtained in Eqs. (59), (63), and (61), respectively, so it shares the instabilities of the grand ensemble.

The projection metric on the $P$ hypersurface is

$$g_{(P)}^{\mu\nu} = g^{\mu\nu} - \epsilon n_{(P)}^\mu n_{(P)}^\nu \quad (89)$$

where

$$n_{(P)}^\mu = \epsilon \frac{g^{\mu\nu}\partial_\nu P(X)}{|g^{\alpha\beta}\partial_\alpha P \partial_\beta P|^{\frac{1}{2}}} \quad (90)$$

and

$$\epsilon = \text{Sign}(g^{\alpha\beta}\partial_\alpha P \partial_\beta P). \quad (91)$$

The components of the induced metric inverse $h_{(P)}^{ij}$ represent the constant pressure fluctuations in the enthalpy $M$ and the angular momentum $J$,

$$h_{(P)}^{11} = \langle (\Delta M)^2 \rangle_P;$$
$$h_{(P)}^{22} = \langle (\Delta J)^2 \rangle_P,$$
$$h_{(P)}^{12} = \langle (\Delta J \Delta M) \rangle_P. \quad (92)$$

Additionally, the components of the projection metric represent the constrained fluctuations in $E$ and $V$,

$$g_{(P)}^{11} = \langle (\Delta E)^2 \rangle_P, \quad g_{(P)}^{22} = \langle (\Delta V)^2 \rangle_P, \text{ etc.}, \quad (93)$$

which bear a simple relation to the enthalpy fluctuations,

$$\langle (\Delta M)^2 \rangle_P = \langle (\Delta E)^2 \rangle_P + P^2 \langle (\Delta V)^2 \rangle_P + 2P\langle \Delta V \Delta E \rangle_P. \quad (94)$$

The fluctuations in $T$ and $\Omega$ are obtained as

$$\langle (\Delta T)^2 \rangle_P = \frac{(l^2 - a^2)\mathcal{N}_{ex}\mathcal{N}_J}{32 l^8 \pi^3 r^2 (a^2 + r^2)^4}, \quad (95)$$

$$\langle (\Delta \Omega)^2 \rangle_P = \frac{(l^2 - a^2)^3 r^2 \mathcal{N}_{ex}}{2 l^8 \pi (a^2 + r^2)^4}. \quad (96)$$

We note that unlike the previous cases, the $T$ fluctuations do not diverge in the $a = 0$ limit.

The scalar curvature in the $P$ ensemble can be obtained directly from the induced metric Eq. (86) or, equivalently, from the grand curvature $R$, Eq. (72), via the Gauss-Codazzi relation, Eq. (40). The scalar curvature was found in [46] as

$$R_P = -\frac{(l^2 - a^2)r^2 \mathcal{P}_4}{l^2 \pi \mathcal{N}^2 \mathcal{N}_{ex}} \quad (97)$$

where $\mathcal{P}_4$ is given in Eq. (A4). Unlike the curvature for the grand ensemble, Eq. (72), and the $J$ ensemble, Eq. (83), the curvature for the $P$ ensemble remains finite in the AdS-Schwarzschild limit $a = 0$. This fits well with the finiteness of the determinant $\mathcal{D}_P$ in Eq. (88) and the temperature fluctuations in Eq. (95). Indeed, in this ensemble the $V$ fluctuations are already subsumed under the fluctuations in enthalpy $M$, so in the limit $a = 0$ the geometry is still two dimensional. We therefore infer from the geometry that the $P$ ensemble remains "feasible" for small $a$ and has a smooth Schwarzschild-AdS limit, unlike the previous two cases discussed. In Fig. 8 we plot contours of $R_P$ in the $r - a$ plane with $l = 1$. It is apparent that $R$ decays along all thermodynamic processes marked by the colored dashed curves.

Just like for the $J$ hypersurface, the extrinsic curvature $K_P$ encodes the instabilities of the ambient state space as well as those of the $P$ hypersurface.

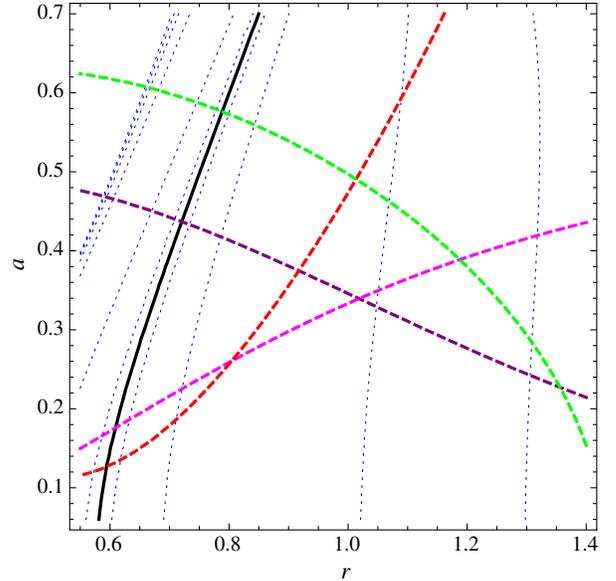

FIG. 8. A plot of contours of the $P$ ensemble scalar curvature $R_P$ as the dotted blue lines in the constant $P$ plane. The thick black curve is the Davies curve where $R_P$ is singular. The dashed red and the magenta curves have respectively $T$ and $\Omega$ constant while the dashed green and purple curves have respectively $M$ and $J$ constant. Notice that now the curvature decays uniformly away from the Davies line. Compare with Fig. 5.

$$K_P = \frac{r^2(l^2 - a^2)(9l^2 r^4(l^2 - r^2) + a^6(5l^2 - r^2) + a^4(11l^4 + 5l^2 r^2) + 3a^2(4l^4 r^2 - 3l^2 r^4 + 3r^6))}{a^2 l^2 (a^2 + r^2)\sqrt{2\pi(r^2 + l^2)}\mathcal{N}\mathcal{N}_{ex}^{1/2}}. \tag{98}$$

The geometry of other ensembles like the fixed volume or the fixed $\Omega$ ensemble can also be similarly investigated. For all of these the state space scalar curvature becomes singular in the AdS-Schwarzschild limit.

## IV. CONCLUSION

In this paper we have developed a prescription for studying the state space geometry in the presence of general constraints. The set of allowed macroscopic fluctuations constitutes a thermodynamic ensemble which corresponds to a hypersurface in the equilibrium state space. The intrinsic scalar curvature of a given ensemble hypersurface, which can also be obtained from the ambient scalar curvature via the Gauss-Codazzi relations, always encodes the thermodynamic instabilities in the given ensemble. The extrinsic geometry of the ensemble hypersurface provides useful information through its normal vector and extrinsic curvature. Thus, while the projection metric on the hypersurface gives the second moments of constrained fluctuations, the trace of the extrinsic curvature is the Lie derivative of the constrained moments along the normal vector. We then applied our prescription to the extended state space geometry for the Kerr-AdS black holes. We studied in detail the geometry of the grand ensemble and two other ensembles, one with fixed angular momentum $J$ and the other with fixed pressure $P$. The grand scalar curvature $R$ is singular at the Davies transition point and also in the Schwarzschild-AdS limit. This indicates that in the grand ensemble of the extended state space, which includes fluctuations in all the extensive state variables— namely, energy, thermodynamic volume, and angular momentum—the limit of zero angular momentum is an instability. In the ensemble with fixed angular momentum, the scalar curvature $R_J$ becomes singular along the spinodal line and hence also at the critical point. This goes to show that, contrary to the prevailing view, thermodynamic curvature encodes critical behavior in Kerr-AdS black holes. Much like the grand ensemble, the $J$ ensemble is also unstable in the AdS-Schwarzschild limit. On the other hand, the scalar curvature for the fixed pressure ensemble $R_P$ is singular along the Davies point but remains finite in the limit $a = 0$. This indicates that the Kerr-AdS black hole in a fixed pressure ensemble, which is the same as the unextended state space, has a well-defined Schwarzschild-AdS limit. Even though in this paper we have focused on black hole systems, our method is quite general and will be useful for studying the thermodynamic geometry of multi-parameter systems.

Finally, we mention that in this work we have focused exclusively on the thermodynamic instabilities and phase transitions in black hole systems, whose geometric treatment is characterized essentially by the Hessian of the entropy. Indeed, there are more general dynamical instabilities of black holes, especially for higher-dimensional objects like black branes and black strings, which might or might not coincide with the thermodynamical ones. In the latter case a more general characterization of instabilities is possible in terms of a canonical energy [82], which could be further interpreted as quantum Fisher information [83]. It would be interesting to investigate the information geometry of the canonical energy and, in particular, implement our proposal for constrained fluctuations in the context of dynamical instabilities.

## ACKNOWLEDGMENTS

We are very grateful to George Ruppeiner, Kalyan Rama, and Gautam Sengupta for a careful reading of the draft of this paper and for providing several useful comments. We also thank Sudipta Mukherji, Tapobrata Sarkar, and Shankhadeep Chakrabortty for useful comments. We thank the referee for bringing Ref. [82] to our attention.

## APPENDIX: FORMULAS

$$\begin{aligned}
\mathcal{P}_1 = &-(a^2 - l^2)r^2(2a^12r^2(l^2 - r^2)(l^2 + r^2)^2 - 27l^4 r^{10}(l^6 - 9l^4 r^2 + 3l^2 r^4 - 3r^6) \\
&+ a^2 l^2 r^8(-203l^8 + 729l^6 r^2 - 423l^4 r^4 + 243l^2 r^6 - 162r^8) \\
&- 3a^6 r^4(151l^{10} - 117l^8 r^2 + 40l^6 r^4 - 252l^4 r^6 + 89l^2 r^8 + 9r^{10}) \\
&+ a^{10}(-36l^{10} - 44l^8 r^2 - 9l^6 r^4 + 81l^4 r^6 + 73l^2 r^8 + 23r^{10}) \\
&- a^4 r^6(395l^{10} - 1061l^8 r^2 + 142l^6 r^4 - 798l^4 r^6 + 351l^2 r^8 + 27r^{10}) \\
&- a^8 r^2(206l^10 + 76l^8 r^2 + 203l^6 r^4 - 71l^4 r^6 + 195l^2 r^8 + 39r^{10}),
\end{aligned} \tag{A1}$$

$$\begin{aligned}
\mathcal{P}_2 = &-(8(-18l^{12}r^{14} + 3a^{16}l^4r^4(l^2+r^2) - 9a^2l^8r^{10}(l^6 + 2l^4r^2 - 22l^2r^4 - 9r^6) \\
&- a^4l^4r^8(22l^{10} - 100l^8r^2 - 512l^6r^4 + 312l^4r^6 + 405l^2r^8 + 81r^{10}) \\
&+ a^{12}l^2(l^{12} - 23l^{10}r^2 - 226l^8r^4 - 518l^6r^6 - 208l^4r^8 - 50l^2r^{10} - 32r^{12}) \\
&+ a^{14}(-l^{12} - 11l^{10}r^2 - 28l^8r^4 + 13l^6r^6 + 11l^4r^8 + l^2r^{10} + 3r^{12}) \\
&+ a^6l^2r^6(149l^{12} + 1241l^{10}r^2 + 2929l^8r^4 + 2119l^6r^6 + 1776l^4r^8 + 1107l^2r^{10} + 243r^{12}) \\
&- a^{10}r^2(-34l^{14} - 140l^{12}r^2 + 362l^{10}r^4 + 1726l^8r^6 + 1307l^6r^8 + 736l^4r^{10} + 390l^2r^{12} + 81r^{14}) \\
&- a^8r^4(-111l^{14} - 715l^{12}r^2 - 704l^{10}r^4 + 1664l^8r^6 + 1848l^6r^8 + 1254l^4r^{10} + 702l^2r^{12} + 162r^{14}))), \quad \text{(A2)}
\end{aligned}$$

$$\begin{aligned}
\mathcal{P}_3 = &-l^4r^8(l^2 - 3r^2)^2 + a^2l^2r^6(9l^6 + 10l^4r^2 - 81l^2r^4 - 18r^6) \\
&+ a^{10}(3l^4r^2 + 4l^2r^4 + r^6) + a^8(-6l^8 - 30l^6r^2 - 35l^4r^4 + 8l^2r^6 + 3r^8) \\
&- a^6r^2(33l^8 + 170l^6r^2 + 274l^4r^4 + 82l^2r^6 + 9r^8) \\
&- a^4r^4(33l^8 + 200l^6r^2 + 404l^4r^4 + 168l^2r^6 + 27r^8), \quad \text{(A3)}
\end{aligned}$$

$$\mathcal{P}_4 = a^6(3l^2 + r^2) + l^2r^2(l^4 - 3l^2r^2 - 54r^4) + 3a^4(2l^4 + 7l^2r^2 - r^4) + 3a^2(l^6 + 7l^4r^2 - 2l^2r^4 - 18r^6). \quad \text{(A4)}$$